\documentclass[12pt]{article}
\usepackage[dvips]{epsfig}
\newcommand{\PY}{{PYTHIA}}
\newcommand{\CO}{{CompHEP}}
\newcommand{\noi}{\noindent}
\newcommand{\eeinto}   {$e^+e^- \longrightarrow \:$}
\newcommand{\geinto}   {$\gamma e \longrightarrow \:$}
\newcommand{\into}   {$\longrightarrow \:$}
\newcommand{\entb}{$e \nu t b \: $}
\newcommand{\gentb}{$\gamma e \longrightarrow  \nu \bar{t} b \: $}
\newcommand{\hww}{$HWW \: $}
\newcommand{\nbbw}{$\nu \,b\, \bar{b}\, W\: $}
\newcommand{\wtb}{$W t b \: $}
\newcommand{\mt}{$m_{t} \: $}
\newcommand{\mh}{$M_{H} \: $}
\newcommand{\ee}{$e^+e^-\: $}
\newcommand{\vtb}{$|V_{tb}| \: $}
\newcommand{\toto}{$t\,\bar{t}\: $}
\newcommand{\tata}{$\tau^+\,\tau^-\: $}
\newcommand{\bb}{$b\,\bar{b}\:\; $}
\newcommand{\cc}{$c\,\bar{c}\:\; $}
\newcommand{\glgl}{$g\,g\: $}
\newcommand{\mm}{$\mu^+\mu^-\:$}
\newcommand{\qq}{$q\,\bar{q}\:\; $}
\newcommand{\nn}{$\nu\bar{\nu}\: $}
\newcommand{\gaga}{$\gamma\gamma\: $}
\newcommand{\ff}{$f\bar{f}\: $}
\newcommand{\lele}{$l\bar{l}\: $}
\newcommand{\ww}{$W^+W^- \: $}

\newcommand{\zz}{$ZZ \: $}
\newcommand{\znull}{$Z^0 \: $}
\newcommand{\hnull}{$H^0 \: $}
\newcommand{\gae}{$\gamma e\: $}
\newcommand{\nene}{$\nu_e\bar{\nu_e}\: $}
\newcommand{\SQRTS}{$\sqrt{s}\:$}
\newcommand{\SQRTSGE}{$\sqrt{s_{\gamma e}}\:$}
\newcommand{\SQRTSEE}{$\sqrt{s_{e^+e^-}}\:$}
\newcommand{\less}{\stackrel{ <}{\sim}}
\newcommand{\gess}{\stackrel{ >}{\sim}}
\newcommand{\Ptmiss}{$p_\perp^{miss}\:$}
\newcommand{\Ptb}{$p_{\perp}^b\:$}
\newcommand{\Ptbb}{$p_{\perp}^{b\bar{b}}\:$}
\newcommand{\Pt}{$p_\perp\:$}
\newcommand{\Ptt}{$p_{\perp}^t\:$}
\newcommand{\Pttot}{$p_{\perp}^{tot}\:$}
\def\PL #1 #2 #3 {Phys. Lett. {\bf#1}              (#3)  #2}
\def\MPL #1 #2 #3 {Mod. Phys. Lett. {\bf#1}        (#3)  #2}
\def\NP #1 #2 #3 {Nucl. Phys. {\bf#1}              (#3)  #2}
\def\PR #1 #2 #3 {Phys. Rev. {\bf#1}               (#3)  #2}
\def\PP #1 #2 #3 {Phys. Rep. {\bf#1}               (#3)  #2}
\def\PRL #1 #2 #3 {Phys. Rev. Lett. {\bf#1}        (#3)  #2}
\def\CPC #1 #2 #3 {Comp. Phys. Commun. {\bf#1}     (#3)  #2}
\def\ANN #1 #2 #3 {Annals of Phys. {\bf#1}         (#3)  #2}
\def\APP #1 #2 #3 {Acta Phys. Pol. {\bf#1}         (#3)  #2}
\def\ZP  #1 #2 #3 {Z. Phys. {\bf#1}                (#3)  #2}
\def\NIM  #1 #2 #3 {Nucl. Instr. and Meth. {\bf#1} (#3)  #2}
 
\newcommand{\BS}{\bigskip}
\newcommand{\VS}{\,\,\,\,}
\newcommand{\SECTION}[1]{\BS\begin{center}{\large\section{\bf #1}}\end{center}}
\newcommand{\SUBSECTION}[1]{\BS{\large\subsection{\bf #1}}}
\newcommand{\SUBSUB}[1]{\BS{\large\subsubsection{#1}}}

\begin{document}

\pagestyle{empty}

\noi DESY 97-123E

\BS\BS

\noi April 1997
\section*{
\vspace{4cm}
\begin{center}
\LARGE{\bf
 The Standard Model Higgs: \\
Discovery Potentials and Branching Fraction Measurements at the NLC
       }\\
\end{center}
}

\vspace{2.5cm}
\large
\begin{center}
 M. Sachwitz, H. J. Schreiber and S. Shichanin${}^*$\\ 
\bigskip \bigskip  

DESY-Institut f\"{u}r Hochenergiephysik, Zeuthen, FRG \\
\end{center}
\vfill
{$*$ On leave of absence from Institute for High Energy
Physics, Protvino, Russia}
\newpage
\pagestyle{plain}
\pagenumbering{arabic}

\section*{Abstract}

We discuss discovery potentials for a 140 GeV Standard Model
Higgs boson produced in~\ee \, collisions at 360 GeV, including
all potential irreducible and reducible background contributions.
In the second part of the study, we estimate the
uncertainties expected for the branching fractions of the Higgs into
\bb, \tata, WW$^{(*)}$ and into $c\bar{c} + gg$ including a realistic
error estimation of the inclusive bremsstrahlung Higgs production
cross section.
\BS\BS\BS

\large
\section{Introduction}

The fundamental particles in the Standard Model (SM) \cite{sm}, 
the gauge bosons, the leptons and quarks, acquire their mass by 
means of the Higgs mechanism \cite{higgs}.
This mechanism implies the existence of a real physical particle, the
Higgs boson, and its discovery  is the most important experiment 
for the standard formulation of the electroweak interactions.

However, in the SM the mass of the Higgs particle is unknown.
A lower bound on the Higgs mass of about \mh $\ge$ 66 GeV has been
established so far from LEP I \cite{leplimit}; this limit can be raised
to $\sim$ 95 GeV in the second phase of LEP with a total energy of
\SQRTS = 192 GeV \cite{cavena}.
Beyond the LEP II range, the multi-TeV $pp$ collider LHC can cover the
entire Higgs mass range up to the SM limit of about 800 GeV
\cite{froidevaux}. 
For \mh \, above $\sim$ 150 GeV, the lepton channel \hnull \into 4$l^{\pm}$
will provide a nearly background free signature,
 while below that value the rare photon decay \hnull \into
\gaga \, is the sole decay channel, established so far.
Due to the overwhelming QCD background, other \hnull \, decay modes
cannot be explored or are difficult to detect. 
In the SM, the intermediate mass range from $\sim$ 110 GeV, derived
from the requirement of vacuum stability for a $\sim$ 180 GeV top
quark mass, to 180-200 GeV is a very attractive region for the
Higgs mass.

Future \ee \, linear colliders operating in the 300 to 500 GeV 
center-of-mass energy range, generically
 denoted as NLC (for Next Linear Collider) in the following, are
the ideal machines to investigate the Higgs sector in this
mass range since it can be easily discovered and all major decay modes
can be explored.
The search for the Higgs boson and the study of
its properties are therefore of primary importance in the physics
program at the NLC.

In order to access the actual capability of such a collider concept 
it is extremely interesting and appealing to apply recently developed
tools, thanks to the effort of several groups, to an analysis of SM Higgs
boson physics. In this study we include
\begin{itemize}
\item{the full matrix elements for 4-fermion final states beyond the
    usual approximation of computing production cross sections 
    of the Higgs and the Z boson times branching fractions into 
    decay products;}
\item{Higgs decay channels with SM decay fractions larger than $\sim$1\%
    treated in a unified
    way;}
\item{initial state QED and beamstrahlung;}
\item{a detector response, with parameters of the detector as designed
    in a series of workshops for the \ee
    linear collider Conceptual Design Report;}
\item{all important reducible background expected to contribute.}
\end{itemize}

In sects.2 and 3~ we present some introduction to linear collider and
detector aspects relevant for Higgs studies. Sect. 4 is devoted to a
description of the event generation procedures. 
We then describe our analysis for the particular case of a cm energy
$\sqrt{s}$ = 360 GeV and a Higgs mass $M_H$ = 140 GeV.
It is shown that the Higgs can be observed with limited
integrated luminosity and the inclusive Higgs cross
section
can be determined with high accuracy, without  assumptions on
the Higgs decay modes.
Once the Higgs is found it is of great
importance to measure the
(relative) Higgs couplings to gauge bosons and fermions. 
We demonstrate the precision with which  the branching fractions 
can be (easily) measured, allowing a potential discrimination of the SM Higgs
from e.g. the lightest CP-even MSSM Higgs over a large range in
tan$\beta$. 
We are particularly concerned with simulations of 
both the signal and background
rates for all important reactions expected to contribute at 360 GeV.
The cm energy is chosen such
that \toto \, pair production as
 background is excluded by definition.

\section{The center-of-mass energy spectrum and luminosity}

The extremely high density to which the beam particles must be focused
to produce sufficient luminosity for particle physics at a NLC results
in a significant interaction rate between the particles of one beam
and the collective electromagnetic fields produced by the particles in
the opposite beam.
A major consequence of this is that the cm energy spectrum at which
\ee \, interaction takes place is not at all monochromatic.
Radiation of photons during the beam-beam collision, the so called 
beamstrahlung,
will result in a \SQRTS-spectrum that depends in detail on the energy
and bunch characteristics of each beam.
In our study we have used, as an example, the beam
parameters for the TESLA design \cite{cdreport}, which are in many
respects typical of
most machine designs studied so far.
In particular, the beamstrahlung simulation used in this study 
includes the effects of
multiple radiation and beam-beam disruption \cite{schulte}. 

\begin{figure*}[h!b]
\begin{center}
\mbox{\epsfxsize=17cm\epsfysize=17.5cm\epsffile{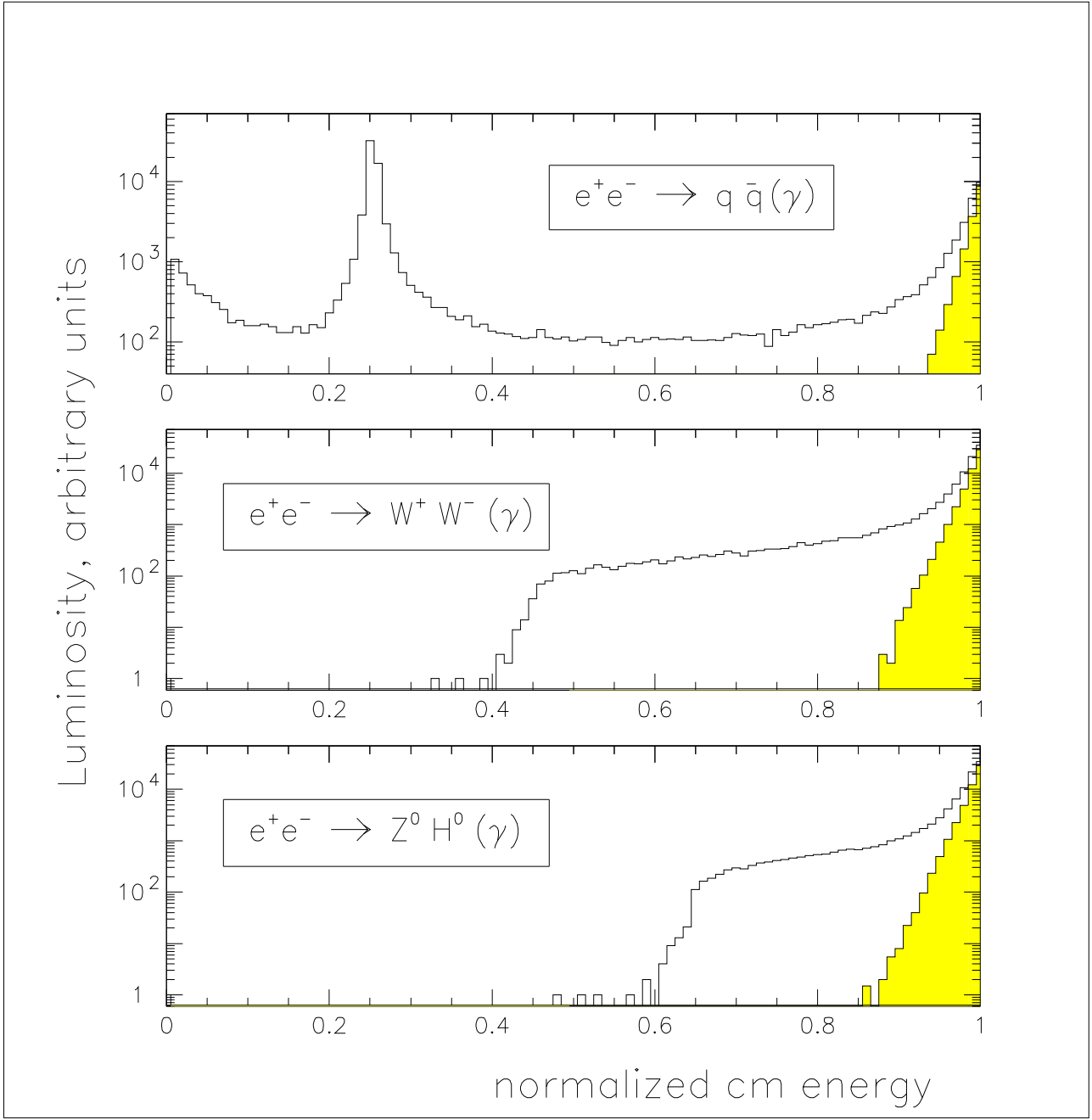}}
\end{center}
\caption[ ]{\sl Luminosity spectra expected after 
convolution with initial state radiation (ISR) 
and beamstrahlung for various reactions. 
The shaded histograms show the energy 
degradation due to the beamstrahlung only. The nominal
 cm energy is 360 GeV.}
\label{fig:isrbeam}
\end{figure*}

Besides beamstrahlung, a further reduction of the initial state energy
occurs due to radiation of photons off the beam particles (ISR).
In our simulation, ISR has been included as suggested in \cite{kuraev} 
and convoluted with the beamstrahlung spectrum. 
The net result of both effects on the cm energy distribution can be seen
in Fig.\ref{fig:isrbeam} for a
few reactions relevant to Higgs studies.

 Here, the hatched histograms reveal the
effect from beamstrahlung alone, for the TESLA design.
In contrast to ISR effects, beamstrahlung energy degradation is,
as expected, reaction-independent.
If the cm energy
cannot be degraded by photon radiation to the $Z^o$ pole,
such as in ZH or WW production, beamstrahlung contributes typically
one-third of the total energy loss during beam-beam collision.
Beam energy spread expected to be $\sim$ 0.1\% has been neglected.

In this study, all  event rates 
have been computed by convoluting the luminosity
spectrum with the relevant cross section evaluated at the reduced cm energy.
Thus, event rates of reactions with an
1/s-cross section dependence are somewhat
enhanced while those with a
logarithmic cross section
s-dependence are only little affected by beamstrahlung and ISR.
Since most of our signal and background rates possesses some
 sensitivity to
the details of the cm energy distribution, we varied
this spectrum within reasonable ranges and found that the
conclusions we draw are not altered.

The luminosity is assumed to be 5 $\cdot 10^{33}$cm$^{-2}$sec$^{-1}$
\cite{cdreport}, so that for two years of running (with
10$^7$~sec per annum)  an accumulated
luminosity of 100 fb$^{-1}$ should be obtained.

\section{Detector response}
 
The properties of a detector that are necessary to successfully carry
out analyses of  the Higgs boson physics 
at the NLC have been developed in
the course of the 'Conceptual Design for an \ee \, Linear Collider and
Detector' of the European high-energy physics community \cite{cdreport}.
The basic components of the detector are a 
vertex detector, a tracker, an electromagnetic and a hadronic 
calorimeter, a muon detector and a luminosity counter.
The parameters of the detector  components  
(resolutions, acceptances and granularities) relevant for this study 
are summarized in Table \ref{tab:1}.
\begin{table}[ht!]
\begin{center}
\begin{tabular}{|l|lll|}
\hline 

&&& \\
Vertexing & $\sigma_{imp} (R\phi)$ & = & 10 $\mu m \oplus 30 \mu m$
$/psin^{3/2} \Theta  $  \\
&&& \\
 & $ |cos \Theta|$ & = & 0.95  \\ [2mm]
&&& \\

\hline 
&&& \\
Tracking & $\sigma_{ p_{\perp}}/{p_\perp}$ & = & 1.5 $\cdot 10^{-4} $
 $ p_{\perp} (GeV)  \oplus 0.01 $ \\ [5mm]
& $|cos \Theta| $ & $\leq$ & 0.95 \\ [5mm]
&  \multicolumn{3}{l|}{radius: $\:$  1.5 m}  \\ [5mm]
&  length: $\:$  5.0 m & \\
&&& \\
\hline
&&& \\
Electromagnet Calorimeter & $\sigma_E/E$ & = & 0.10 /
$\sqrt{E(GeV)} \oplus 0.01 $ \\ [5mm]
& \multicolumn{3}{l|}{cell size: \hspace{0.5em} $0.7^o \times 0.7^o$ } \\ [5mm]
& $|cos \Theta|$ & $\leq$ & 0.98481 $\; (\hat{=} \hspace{0.2em}
10^o ... 170^o)$ \\ [2mm]
&&& \\

\hline
&&& \\
Hadronic calorimeter & $\sigma_E/E$ & = & 0.50 /
$\sqrt{E(GeV)} \oplus 0.04 $ \\ [5mm]
& \multicolumn{3}{l|}{cell size: \hspace{0.5em}  $2^o \times 2^o$} \\ [5mm]
& $|cos \Theta|$ & $\leq$ & 0.98481 $\; (\hat{=} \hspace{0.2em}
 10^o ... 170^o)$ \\ [2mm]
&&& \\
\hline

&&& \\
Muon detector & \multicolumn{3}{l|}{Fe yoke instrumented as tail catcher} \\
& \multicolumn{3}{l|}{and muon tracker} \\
&&& \\

\hline
&&& \\
Luminosity counter & $ \sigma_E/E$ & = & 0.12 / $\sqrt{E(GeV)}$
$\oplus 0.02 $ \\ [5mm]
& \multicolumn{3}{l|}{cell size: \hspace{0.5em} $2^o \times 2^o$}  \\ [5mm]
& \multicolumn{3}{l|}{$4^o < \Theta  < 8^o;$  \hspace{1cm} 
$172^o < \Theta < 176^o$}   \\ [2mm]
&&& \\
\hline
\end{tabular}
\end{center}
\caption[ ] {\sl Detector parameters. The symbol $\oplus$ means
  quadratic sum.}
\label{tab:1}
\end{table}
As can be seen, for the detector properties we assume no extension of 
any technology beyond what exists or is feasible in the near future,
although we can anticipate that improvements in the 
techniques and technologies used in design and construction 
of detector components will occur.
In this analysis the emphasis is addressed to the tracker, the vertex 
detector and the calorimeters.

The detector response is obtained by modifying
 the generated particle momenta, energies, directions
 and nature in the following manner.
At first, particles within the acceptance regions are 
divided into isolated particles and energetic clusters.
Particles are called isolated if they have a 'large' distance 
to any other when entering the electromagnetic calorimeter, 
otherwise they are combined to clusters.
The isolation criteria are based on the cell size of the calorimeters.
For clusters, the electromagnetic and hadronic energy components undergo
separate smearing by a Gaussian according
to the parameters of Table \ref{tab:1} and are then combined 
to the total cluster energy.
Their directions are also Gaussian-modified according to the
granularities. Note
that for charged isolated particles, the smeared momentum from the 
tracker is used except in the case when a better energy value 
from the calorimeter exists.
We apply for the transverse momentum smearing the formula

\begin{displaymath}
\frac{\sigma_{ p_{\perp}}}{p_\perp} = A p_\perp \oplus 1\%
\end{displaymath}

with A = 7 $\cdot$ 10$^{-5}$. This resolution parameter
 is expected from the combined vertex detector and the central
tracker together with a constraint from the main \ee \, vertex.
The tracker resolution scaling inversely proportional with the 
square of the projected track length is accounted for in the program.
Azimuthal and polar angles of charged particles are smeared with 
standard deviations of 1 mrad and 2 mrad, respectively. 
Isolated neutral hadrons (photons) are modified according to the resolution parameters of the hadron (electromagnetic) calorimeter.
Charged particles are required to have a minimum
transverse momentum of 200 MeV, while an
electromagnetic (hadronic) calorimeter 
entry should have at least a deposited energy of 100 (300) MeV;
otherwise they are removed.
Electrons (muons) are misinterpreted as charged pions with 1\% (2\%) 
probability, and  muons should exceed an energy of  3 GeV.
A solenoidal magnetic field of 3 Tesla is assumed.

The relatively  long lifetime of the $b$-quark,
expected preferentially to be produced from Higgs 
boson decays, gives rise 
to decay vertices displaced from the 
primary \ee \, vertex or, equivalently, to tracks with large impact parameters.
We included these decays by simulating the performance of 
a vertex detector and making use of the
impact parameter information in the plane perpendicular to the beam direction.
For each track projection we compute its 
distance of closest approach to the primary vertex, b, and assume that the 
measurement error on b can be parameterized as 
\begin{displaymath}
\sigma_b = \sqrt{A^2 + [B/(p \sin^{3/2}\Theta)]^2} \: ,
\end{displaymath}
where the parameters A = 10 $\mu$m and B = 30 $\mu$m reflect the 
intrinsic resolution of the vertex detector and the effect of 
multiple scattering of particles traversing the beam pipe and the detector.
A charged particle is defined to
be a 'high impact parameter track' if it is within the vertex 
detector acceptance region and has a value of 
b$_{norm}$ = b/$\sigma_b$ greater than two (or three).
An upper limit of the impact parameter has also been 
introduced to suppress contamination
from $K_S$ and $\Lambda$ decays.
Our vertex detector simulation is kept simple and generic  
so that the success of the analysis does not depend on specific 
details of the device.
As soon as a detailed design of the vertex detector exists, 
dedicated Monte Carlo simulations will be mandatory.

\section{Event generation}
 
We have assumed that the Standard Model with three generations of 
quarks and leptons is a correct description of nature and the
Higgs particle couples to bosons and fermions as predicted by the SM
model. 

The reactions considered in this note are listed in Table \ref{tab:2}.
\begin{table}[h!t]
\begin{center}
\begin{tabular}{|llc|c|c|l|}
\hline\noalign{\smallskip}
 & & & & signal  & \\
\multicolumn{2}{|c}{Reaction} & & cross section, fb& cross section, fb&
\multicolumn{1}{|c|}{comments } \\
\hline\noalign{\smallskip}
 & & & & & \\
$e^+e^-$ & $\rightarrow \mu^+\mu^-q\bar{q}$ &(1) & 50.0 & 2.85& see
text\\
 & & & & & \\
$e^+e^-$ & $\rightarrow e^+e^-q\bar{q}$ &(2) & 4450 & 4.62 & see
text\\
 & & & & & \\
$e^+e^-$ & $\rightarrow \tau^+\tau^-q\bar{q}$ &(3) & 51.9 & 2.96
&$H^0\rightarrow\tau^+\tau^-$ and\\
 & & & & & $Z^0\rightarrow q\bar{q}$ only\\
$e^+e^-$ & $\rightarrow \nu\bar{\nu} q\bar{q}$ &(4) & 347 & 33.0& see
text\\
 & & & & & \\
$e^+e^-$ & $\rightarrow 2q 2\bar{q}$ &(5) &434 &55.6 & see text\\
 & & & & &  \\  \hline
 & & & & &  \\
$e^+e^-$ & $\rightarrow W^+W^-$ &(6) & 11200 & - &\\
 & & & & & \\
$e^+e^-$ & $\rightarrow e\nu W$ &(7) &3650 & - & $WW$ excluded\\
 & & & & & \\
$e^+e^-$ & $\rightarrow q\bar{q}(\gamma)$ &(8) &33900 & - &\\ & & & & & \\
 & & & & & \\  \hline
 & & & & & \\
$e^+e^-$ & $\rightarrow H Z $ &(9)& - &18.5 & \\
 & $\rightarrow WW^*Z$ & & & & \\
 & $\rightarrow 3q 3\bar{q}$ & &  & & \\
 & & & & & \\  \hline
 & & & & & \\
$e^+e^-$ & $\rightarrow WWZ $ &(10) &4.21 & - &\\
 & $\rightarrow  3q 3\bar{q}$& & & & \\
 & & & & & \\
$e^+e^-$ & $\rightarrow ZZZ$ &(11) & 0.31& - &\\
& $\rightarrow  3q 3\bar{q}$ & & & & \\
 & & & & & \\  \hline
& & & & & \\
$e^+e^-$ & $\rightarrow t \bar{t}$ &(12) & - & - &\\
& & & &  & \\ \hline
& & & & & \\
$e^+e^-$ & $\rightarrow \nu\bar{\nu} gg$ &(13) & - &2.36 & \\
& & & &  & \\ \hline
& & & & & \\
$e^+e^-$ & $\rightarrow q\bar{q} gg$ &(14) & - &4.20 & \\
& & & &  & \\
\hline\noalign{\smallskip}
\end{tabular}
\end{center}
\caption[ ]{\sl Reactions involved in this study and their total and
  signal  cross sections at  \SQRTS  = 360 GeV and
  $M_H$ = 140 GeV. ISR and beamstrahlung are included. We assume a top
  mass being somewhat above 180 GeV. }
\label{tab:2}
\end{table}

For the
2-to-4 body reactions (1) - (5), the full matrix elements are used
for event generation,
beyond the usual approximation of computing production cross sections
times  branching fractions. In this way,
contributions
of non-leading diagrams, interferences, correct spin structures and
non-zero fermion masses are taken into account. The calculation
consists of two main steps. The generation of Feynman diagrams, 
the analytical expressions for the matrix elements squared and the
corresponding optimized Fortran code have been obtained 
by means of the computer package CompHEP \cite{comphep}. The
integration over phase space and the generation of the event flow
have been done with the help of the adaptive Monte Carlo package
BASES/SPRING \cite{bases}. It is worthwhile to point out that because of
the complicated phase space structure of  4-fermion final 
states and the occurrence of singularities, an appropriate 
choice of variables and smoothing of singularities were 
mandatory. Today's version of CompHEP offers all features 
needed to overcome these problems.   

So far, the reaction \ee $\rightarrow \mu^+ \mu^- b\bar{b}$~ has
been studied in e.g. \cite{boos1}. It has been found
that this reaction has the
cleanest signature for the Higgs but the event rate is rather
small. At least six times more Higgs events
are expected in the reaction \eeinto
\nn \bb, for which  complete
tree-level calculations can be found in \cite{boos2}. Here, some extra
contribution comes from the fusion process 
\eeinto \nn \hnull. The reaction
\ee $\rightarrow$ \ee $b \bar{b}$~ turned out to be
the most complicated one from the calculational point of view
\cite{boos3}, and the extraction of the Higgs out of a background two
orders of magnitude larger requires well designed cuts. Higgs
production and its detection potential have also been considered in
the reaction \ee $\rightarrow b \bar{b}$ + 2 jets \cite{ballestr}.

These studies restricted to b-quarks in the final state
have now been extended to include
the quark states u, d, s, c
 and b. In addition, reaction (3), 
 \ee $\rightarrow \tau^+ \tau^- q \bar{q}$, is included in the analysis. Its
tree-level calculation is performed in analogy with that of reaction
(1), \ee $\rightarrow \mu^+ \mu^- q\bar{q}$, with the substitution
$m_{\mu} \rightarrow m_{\tau}$ and the restrictions to 
$Z^0 \rightarrow  q \bar{q}$ and $H^0 \rightarrow \tau^+\tau^-$ decays.

In order to avoid unnecessary large  event samples the
following conditions were applied during event generation:

\begin{itemize}
\item for reactions (1)-(3):  M(\qq)$>$ 50 GeV, $|$cos $\Theta$
$(q,\bar{q})| <$ 0.993 and M(\lele) $>$  30 GeV; 
 \item for reaction (4): either M(\qq)$>$ 50 GeV and $|$cos $\Theta$
$(q,\bar{q})|<$ 0.993 or M(\qq)$>$ 100 GeV and $|$cos $\Theta$
($e^+) |<$ 0.9962 and $|$cos $\Theta$ ($e^-) |<$ 0.9962;
 \item for reaction (5):  M(\qq)$>$ 10 GeV, $|$cos
$\Theta(q,\bar{q})| <$ 0.993, $|$cos $\Theta (q)| <$ 
0.993 and $|$cos $\Theta (\bar{q})| <$ 0.993.
\end{itemize}
\vspace{1cm}

At cm energies 300-500 GeV, high $p_t$-background contributions 
are theoretically well understood and accurately
calculable. Most of the background expected is due
to hard electroweak and QCD processes. They are also listed in Table 2,
together with their cross sections. As can be seen, these
reactions  contribute more than 4 million events to the data
volume for an
integrated luminosity of 100 $fb^{-1}$.  Such an enormous
data sample might obscure and/or mimic a Higgs signal. Hence, we
include these channels in our simulation procedures at
the same level as the signal reactions in order to make a
signal-to-background analysis as meaningful as possible.

A Higgs boson with a mass of 140 GeV decays
in almost 50 \% of the cases into $WW^*$. The knowledge  of this
branching fraction is of importance because it helps greatly 
to establish the nature of the Higgs and, together with the cross
section of the WW-fusion reaction, to determine the total width of
the Higgs. In order to ensure large
statistics for the \hnull \into WW* decay, the $W \rightarrow q\bar{q}$
and  $Z \rightarrow q\bar{q}$ 
decays are considered in this paper. It would be desirable to
embed the signal channel \ee \into \znull\hnull \into (\qq)(WW*)
 in a general 
2-to-6 body event generator, so that besides the signal  irreducible
background and interferences could be also addressed.
Such an event generator
is at present not available. Therefore, the signal channel has been
processed in the usual approximation
of computing production cross sections times the \hnull \into WW* branching
fraction, as indicated in Table 2.
The SM 6-jet 
processes, \ee $\rightarrow WW Z$ and $ZZZ$, are
expected to be the only significant sources of background; they were
generated in the same manner as the Higgs channel, with cross
sections as listed in Table 2.

The \hnull \into \cc + gg  decay occurs in a few percent of
the cases. We try to isolate this decay mode using the 2-jet missing
energy and the 4-jet reactions (4), (5), (13) and (14). We would like
to point out that reaction (13) is free of tree-level background,
whereas the \qq gg  background is already accounted for by gluon
radiation in reaction (8).

Reactions (6),~(8),~(13) and~(14) have been generated by means of the
package PYTHIA 5.7 \cite{pythia} ~including the
beamstrahlung effects \cite{ohl}, while for reactions (7) and (9)-(11)
~\CO ~has been used.
For all reactions considered,
the JETSET 7.4 package \cite{pythia} has been used for quark and gluon
fragmentation and unstable particle decays. The events
generated  are fed into a fast detector simulation program as
described in sect. 3.

One should emphasize that, to a good approximation, dedicated
calculations of the electroweak part of the reactions
considered have achieved very high accuracy but that
relatively large uncertainties exist in the perturbative parton shower
or non-perturbative hadronization procedures. This rises the question
of the reliability of such calculations for the study of hadronic and
mixed hadronic-lepton final states. In this paper we
 adopted a pragmatic solution: we used a standard interface program 
for parton-shower and hadronization procedures and rely on its
usefulness for our application.

\section{Higgs Discovery Potential}

Basically, the Higgs can be produced by the
bremsstrahlung process\cite{brems} 

\setcounter{equation}{14}
\begin{equation}
e^+e^- \longrightarrow Z^* \longrightarrow Z^0H^0
\end{equation}

\noi or by the fusion of $WW$ and $ZZ$ bosons\cite{fusion}

\begin{eqnarray}
e^+e^- & \longrightarrow& \nu \overline{\nu} H^0\\
e^+e^- & \longrightarrow& e^+e^- H^0 \,.
\end{eqnarray}

At \SQRTS = 360 GeV and $M_H$ = 140 GeV, the bremsstrahlung process is
about four times more important than the fusion reactions. This is in 
good approximation also true if ISR and beamstrahlung effects are
included. Therefore we restrict ourselves to the bremsstrahlung process
in most of the cases and try to select it from all the background expected
to contribute.

The bremsstrahlung process admits two main strategies for the Higgs search:
\begin{itemize}
\item calculation of the mass recoiling against the $Z^0$, most 
conveniently applied in the $Z^0$ \into \ee \,and \mm \,decay modes.
This method has the unique feature of being 
independent on assumptions about the Higgs decay modes;
\item direct reconstruction of the invariant mass of the Higgs 
decay products; here the decays \hnull \into \bb, WW, 
$\tau\bar{\tau}$ \,and light \qq \, or \glgl \, are appropriate,
with the \bb \,decay mode being the most effective
one for the \hnull \,discovery, and, if possible, the fusion processes
are included.

\end{itemize}

In the past, various \hnull \,analyses 
have been developed for 
practically all possible \hnull \,and \znull \,decays, 
see e.g. \cite{pgw,janot,bark}.
Here, we follow the guidelines of these studies, but have 
either improved the selection criteria or added further 
restrictions or impose energy-momentum as well as 
$M(l^+l^-) = M(Z^0)$ constraints, when appropriate,
in order to achieve better experimental resolution.
If for a particular \hnull \,and \znull \,decay a set of cuts and
constraints had been established, all reactions of Table 2 were also 
processed under these conditions.
Their contributions to either the recoil mass or the appropriate 
invariant mass were added so that the Higgs signal over the
corresponding  background can be explored.

\begin{figure*}[h!b]
\begin{flushright}
\mbox{\epsfxsize=17cm\epsfysize=17.5cm\epsffile{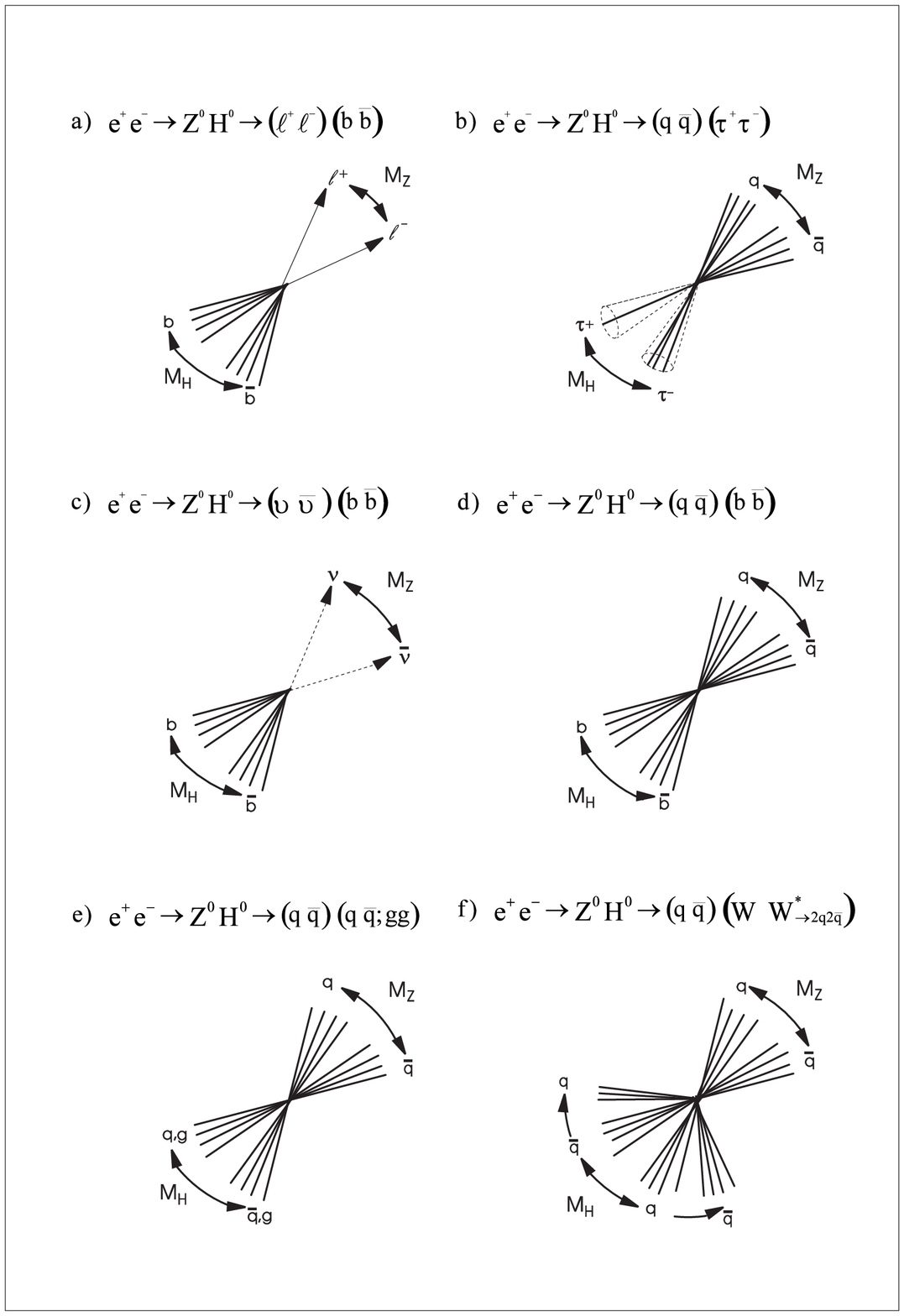}}
\end{flushright}
\caption[ ]{\sl Higgs signal topologies expected for the reactions studied in this paper.}
\label{fig:schrei5}
\end{figure*}

Higgs signal topologies expected are shown in Fig.\ref{fig:schrei5}.
Typically the associated  \ee \into \znull \hnull production
\,is followed by 
the decay of the \znull \into \lele (10\%), \nn (20\%), \qq(70\%)
and the decay of the \hnull \,mostly into \bb, $W W^*$ and
occasionally into \tata \, or into light quarks \qq \,and
two gluons. Only $W \rightarrow q\bar{q}$ decays are considered here.

For each of the reactions  we apply so-called standard and improved cuts
so that the background is removed to a negligible (or acceptable) level while
the signal is retained to a large extent. Standard cuts
are mainly introduced to suppress non-high $p_\perp$, low-multiplicity
events whereas the improved cuts are chosen to remove
effectively WW and $q\bar{q}$ pairs as well as $\;e\nu W$, WWZ and
ZZZ background contributions and possible reflections of any of the signal
channels (1) - (5), (9) onto that one under study. The numbers of signal
and background  events, S and B, are then counted  in a window  around the
Higgs mass   in order to determine the statistical
significance $S/\sqrt{B}$ of the signal.
 This quantity is studied as a function of the accumulated luminosity and, if
$S/\sqrt{B}$ is typically 5 or larger, Higgs detection should be
promtly feasible either in the
recoil mass or in the invariant mass of its decay products.
 It is worthwhile to note that the
proposed criteria are robust and relatively simple, and they are
not optimized such that significant improvements are expected to be possible
once the detailed detector behaviour is known. Our concern today is to
demonstrate the
potential of Higgs discovery at the NLC
 by the analysis methods proposed here.

\subsection{The leptonic channel: \ee \into \znull\hnull \into 
(\mm ;\ee)(\bb)}

According to the topology in Fig.2a, the Higgs can be detected either
by 
calculating the recoil mass, $M^2_{rec} =  s - 2 \sqrt{s} E_Z +
M_Z^2$, once two-opposite charged leptons with invariant mass close to
the \znull $\;$ mass are identified, or by
investigation of the hadronic 2-jet mass.

The set of cuts we have adapted to analyze this channel consists of

\begin{itemize}
\item [a)] the total transverse energy of the event has to be larger 
than 30 GeV but less than 250 GeV; 

\item [b)] the total momentum along the beam is restricted to be
  within the range $\pm$  120 GeV;

\item [c)] the visible energy of the event should be larger than 230 GeV;
\item [d)] the total number of tracks exceeds 14;
\item [e)] $M$(\mm) = $M_Z \pm$ 10 GeV, $\:$ respectively,
 $M$(\ee) = $M_Z \pm$ 6
  GeV,  \\
with $|cos\Theta_Z| <$ 0.8;

\item [f)] if the $Z^0$ decays into two muons,
the energy of isolated 
electrons has to be  smaller than 25 GeV, whereas for  $Z^0$
\into \ee \, decays, the electron energy
 is required to be between 10 and 140 GeV;

\item [g)] if the hadronic system contains two jets\footnote{The
    LUCLUS\cite{pythia} jet finding algorithm has been used in
    this study.} 
with an opening angle larger than $40^o$, each jet has to have an energy
larger than 10 GeV, a number of particles of at least
 5 and $|cos \Theta_{jet}| <$ 0.8;

\item [h)] the number of tracks with large impact parameter, 
$b_{norm} \ge$2, has to
  be larger or equal than 3; here, the $Z^0$ deay products are
not acounted for.
\end{itemize}

Conditions a) - d) are very useful to reject most of low \Pt,
low-multiplicity background; they do not remove signal
events. Constraint f) aims to suppress (irreducible) background from
the reaction \ee \into \ee \qq. 
The remaining criteria are  chosen to isolate
\hnull ~events from the overwhelming WW and \qq($\gamma$) background, to
ensure good particle measurability and to enrich \hnull \into \bb
decays. In the case of the recoil mass, we applied in addition the
constraint M(\lele) = M(\znull) to improve the $S/\sqrt{B}$
ratio.

\begin{figure*}[h!t]
\begin{center}
\mbox{\epsfxsize=17cm\epsfysize=17.5cm\epsffile{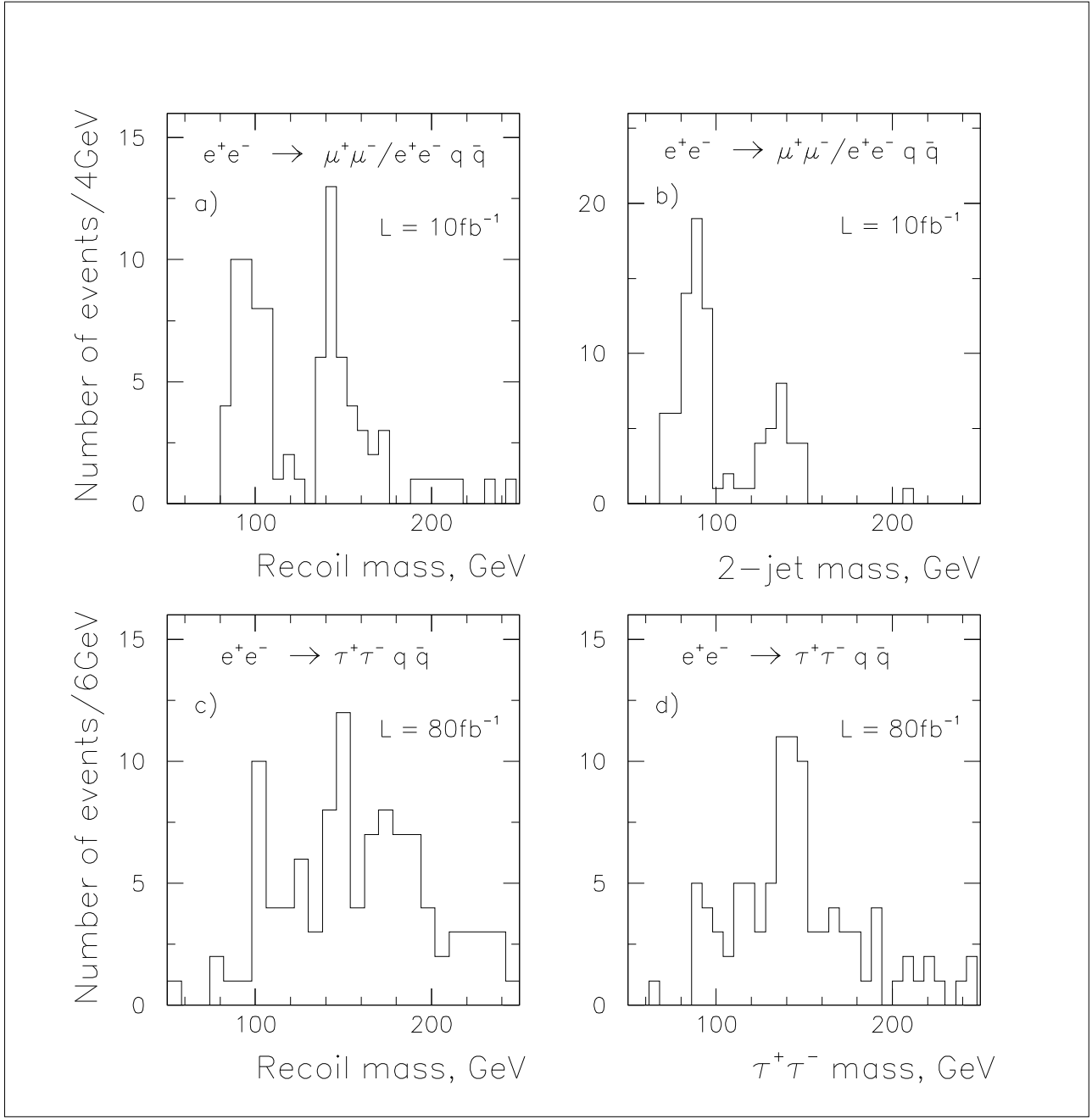}}
\end{center}
\caption[ ]{\sl Recoil and 2-jet (\tata) mass distributions for the reactions \ee \into \mm/\ee\qq \,and \tata\qq \,with integrated luminosities as indicated.
Background contributions are included.}
\label{fig:eemmtau}
\end{figure*}

After all cuts applied we are left with 38\% of signal events for
further analyses. The determination of the integrated luminosity needed to
observe the Higgs with $S/\sqrt{B} \ge$ 5 results, for
example, in the histograms of Figs.\ref{fig:eemmtau}a and b. Here, the recoil mass as
well as the 2-jet mass are shown for 10 fb$^{-1}$ accumulated
luminosity, with
either 21 or 23 \hnull events 
in the range $M_H \pm ^{20}_{6}$ GeV,
respectively,  $M_H \pm ^{10}_{15}$ GeV over a very small background
(of about 2 events). 

A non-gaussian behaviour of the reconstructed masses is observed:
the recoil mass has a tail on 
the large side due to ISR and beamstrahlung energy
degradation, while the 2-jet mass indicates a tail to the lower side
due to imperfect detector acceptances and neutrinos in the final
state.

\subsection{The tauon channel: \ee \into \znull\hnull \into (\qq)(\tata)}  

Since one of our interests is to measure the branching fraction BF (H
\into \tata) from the 4-body final state \tata \qq
only events with
 \znull \into \qq and \hnull \into \tata decays are selected. This
event rate is about two times more abundant than the complementary rate with \znull
\into \tata and \hnull \into \qq decays. Accordingly,
 Higgs detection is
possible  by inspection of the \tata mass and, once the
\znull \into \qq \, decay is established, of the mass recoiling against
the \znull. The final state of interest here is formed by two
isolated energetic tau's and two jets as indicated in Fig. 2b.

In order to distinguish a possible signal from all background
contributions we apply at first the following cuts:

\begin{itemize}
\item [a)] the total transverse energy of the event 
should be in the range 30 GeV to 240
  GeV;
\item [b)] the total momentum along the beam is restricted to be within
  $\pm$140 GeV;
\item [c)] the visible energy of the event is larger than 150 GeV but below
  330 GeV, to allow for missing energies by neutrinos from $\tau$ decays;
\item[d)] the number of charged particles should exceed 8;
\item [e)] isolated electrons with a polar angle between
15$^0$ and 165$^0$ should have an energy less than 125 GeV.
\end{itemize}

\noindent
The two $\tau$'s are selected as the two most isolated oppositely
charged particles\footnote{For simplicity, only 1-prong $\tau$ decays
  are considered.} with
\begin{itemize}
\item [f)] a transverse momentum of at least 3.5 GeV;
\item [g)] no other charged particles in a cone of 25$^0$ around their direction;
\item [h)] not identified as an \ee \,or \mm \,pair;
\item [i)] all neutrals within a cone of 12.5$^0$ around the charged
particle direction are included.
\end{itemize}

The particles selected by criteria f) - i) are
then subtracted from the event and the remainder 
is considered to be a 2-jet system with an invariant  mass within 
15 GeV of the $Z$ mass.
Since we know the cm energy and momentum, the initial $\tau$ and jet
energies are recomputed by a fit; otherwise no 
\hnull signal would  be visible in the recoil and the
\tata  mass spectra.

In order to ensure good event containment, reasonable track
reconstruction capability, well-defined \znull \into 2-jet decays
and suppression of ZZ events
we require additionally

\begin{itemize}
\item[j)] the $\chi^2$, \\ [2mm]
\begin{math}
\chi^2 = \frac{(M_Z-M_{jj})^2}{\sigma^2_{M_{jj}}} \:+\:
\frac{(M_H-M_{\tau\tau})^2}{\sigma^2_{M_{\tau\tau}}} ,
\end{math} \\ [2mm]
is less than 25. The $\sigma$'s ~are the corresponding mass resolutions;
\item [k)] $|cos\Theta_Z| <$ 0.8;
\item [l)]the two jets found should have an opening angle larger than
40$^o$ and, for each jet, it is demanded to have an energy larger 
than 10 GeV, $|cos \Theta_{jet}| <$ 0.8 and to involve at least 5 particles.
\end{itemize}

Some 15 \% of the signal events survive all cuts. However,
non-negligible background remains due to insufficient~ ZH
particle
association from the \tata \qq ~final state and due to WW
pair and $2q2 \bar{q}\; $ background production. In addition, the small
H \into $\tau\tau$ branching fraction of 4.5 \% together with degraded mass
resolutions makes a Higgs discovery in the $\tau\tau$\qq channel
not easy. Large statistics and well-educated selection procedures are
prerequisites to observe a convincing signal. The results of our
analysis for 80 fb$^{-1}$ integrated luminosity
 are shown in Figs. 3c
and d. In the \tata mass distribution a clear signal of $\sim$ 22
events over  10 background events is expected in the mass range 
$M_H \pm$ 12 GeV, whereas the recoil mass seems to be less useful for finding
the Higgs.

\subsection{The missing energy channels: \ee \into \znull\hnull \into
  (\nn)(\bb) \\
\hspace{17cm} and \ee \into \nn \hnull \into \nn (\bb)}

The missing energy channel is of great interest because of its potentially
large discovery potential for
the Higgs  with limited integrated luminosity, due to the existence
of two signal diagrams and the large branching fractions
 BF(\znull \into \nn) and BF(\hnull \into \bb).

The topology expected is an (acoplanar)
pair of jets accompanied by large missing energy.
The difference between the two \hnull 
production mechanisms consists basically
in the constraint M(\nn) = M$_Z$, which might be applied
 to select the bremsstrahlung process, see Fig.2c.
The isolation criteria we have used to select the signal
from background processes are:

\begin{itemize}
\item[a)] the total transverse energy of the event should be 
larger than 10 GeV but less than
  170 GeV;
\item[b)] the total momentum along the beam line is within the range
  $\pm$ 140 GeV;
\item[c)] the visible event energy is larger than 50 GeV and less than
  210 GeV, to allow for large missing energy;
\item[d)] the number of charged particles is larger than 6;
\item[e)] isolated electrons have an energy smaller than 50 GeV and
  their polar angle should be between $20^o$ and $160^o$;
\item[f)] the invisible mass recoiling against the 2-jet system has to
  be within 60 and 250 GeV;
\item[g)] the hadronic system involves two jets with an opening angle
  of at least $30^o$;
\item[h)] each jet should have an energy larger than 10 GeV, involves at least
  5 particles and has a polar angle with $|cos\Theta_{jet}| <$ 0.8;
\item[i)] the number of tracks with large impact parameter,
  b$_{norm} > 3$, is required to be 5 or larger.
\end{itemize}

Cuts a) - e) reduce background to a large extent; they do not
remove  signal events significantly. The other criteria 
should accumulate well-defined b-quark jets. Note that we have
strengthened the definition for 'large impact parameter' track
with the result that remaining $W^+W^-$, e$\nu$W, \qq$(\gamma)$ and
\ee\qq \, background rates are reduced to a very small level.

\begin{figure*}[h!t]
\begin{center}
\mbox{\epsfxsize=17cm\epsfysize=17.5cm\epsffile{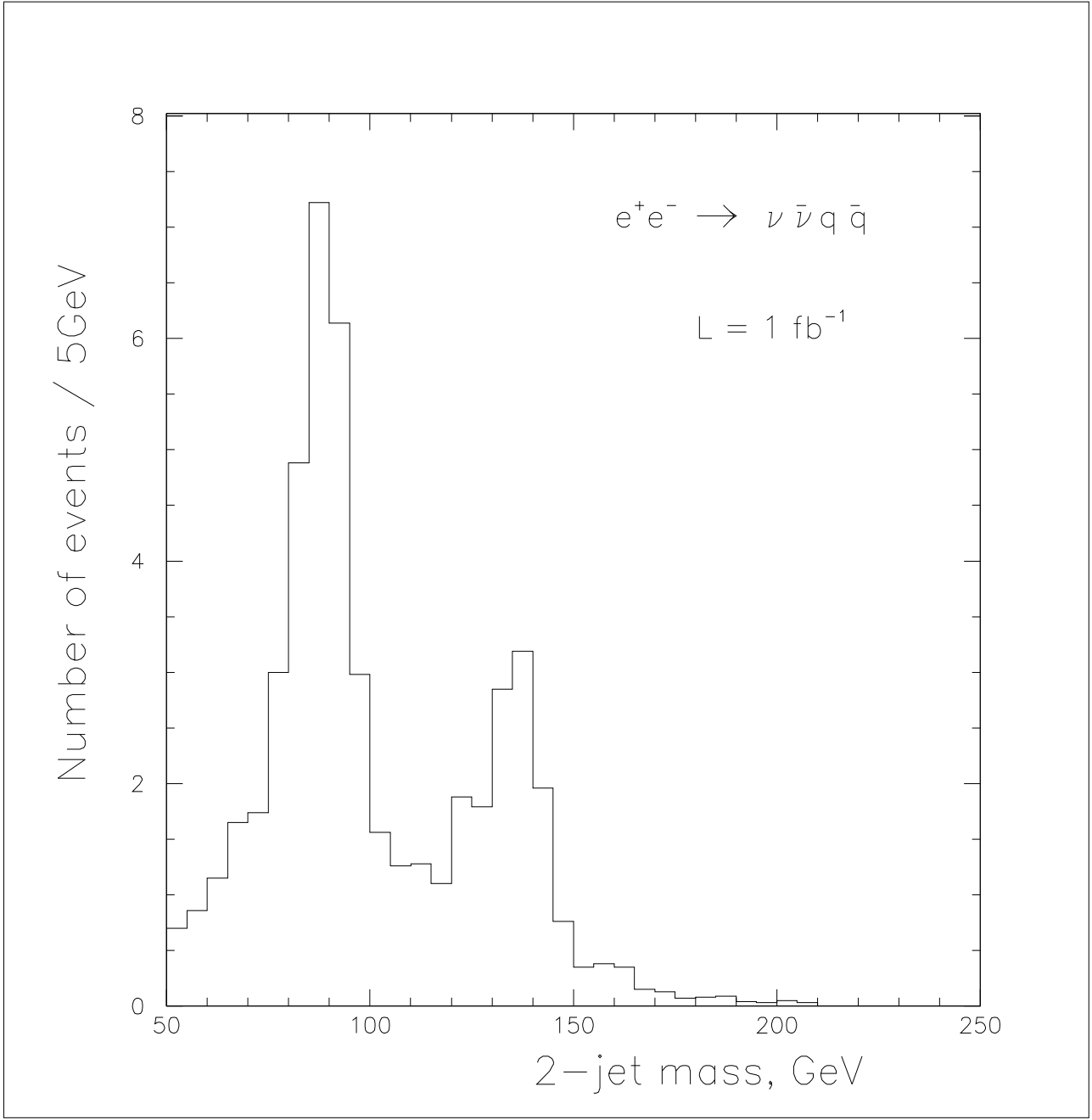}}
\end{center}
\caption[ ]{\sl 2-jet  mass distributions of the reaction \ee \into \nn\qq \,for an integrated luminosity of 1 fb$^{-1}$.
Background contributions are included.}
\label{fig:neutrinos}
\end{figure*}

After application of all criteria,
36 \% of the signal events survive which,
together with the relatively large cross sections  for the
bremsstrahlung and fusion processes times the corresponding \znull ~and
\hnull ~branching fractions, allows to reduce the integrated
luminosity to 1 $fb^{-1}$ to observe 8-9 signal events over about 3
background events within the mass range 120 to 145 GeV, as can be seen in
Fig.\ref{fig:neutrinos}. Due to fluctuations expected within
such small event samples we conclude that 1.5 to 2 $fb^{-1}$ of
accumulated luminosity should be sufficient to observe a significant
signal.
This result makes the \nn\qq channel
very attractive for a quick Higgs search in \ee
collisions at cm energies of $\sim$ 300 to 360 GeV.

\subsection{The 4-jet channel: \ee \into \znull\hnull \into
  (\qq)(\bb)}
The topology of this channel (Fig. 1d) involves 4 jets from the
\znull \into \qq and \hnull \into \bb $\:$ decays. Again,
two possibilities for the Higgs search exist; the
recoil mass and the \bb  invariant mass can be searched for a signal.
Since numerous backgrounds are expected to contribute to this
topology a more sophisticated analysis has to be developed.
We have come up with the following selection criteria:

\begin{itemize}
\item[a)] the total transverse energy of the event 
should be larger than 30 GeV and
  smaller than 270 GeV;
\item[b)] the total momentum in beam direction has to be in the range
  $\pm$ 120 GeV;
\item[c)] the visible energy exceeds 180 GeV;
\item[d)] the number of charged particles should be larger than 24;
\item[e)] isolated electrons should have a polar angle between 20$^o$
  and 160$^o$ and an energy smaller than 60 GeV;
\item[f)] the event is accepted if the number of jets found equals 4
  and the opening angle between any pair is larger than 20$^o$;
\item[g)] from all possible Z and H jet pairing only the
  combination with smallest $\chi^2$, \\
\begin{math}
\chi^2 = \frac{(M_Z-M_{ij})^2}{\sigma^2_{M_{ij}}} \:+\:
\frac{(M_H-M_{kl})^2}{\sigma^2_{M_{kl}}} ,
\end{math} \\
is selected. M$_{ij}$ is the invariant mass of the jet-pair (i,j) and
$\sigma_{M_{ij}}$ is its mass resolution;
\item[h)] the four jet energies are recomputed using energy-momentum
  conservation, while keeping their directions at measured values; 
\item[i)] the polar angle of the 2-jet system consistent with the
  \znull is restricted to $|cos \Theta_Z| < 0.8$;
\item[j)] each jet should have an energy larger than 10 GeV,  $|cos
  \Theta_{jet}| < 0.8$ and contains at least 5 particles;
\item[k)] the number of tracks with significant impact parameter,
  $b_{norm} > 3$, is 5 or larger.
\end{itemize}

\begin{figure*}[h!t]
\begin{center}
\mbox{\epsfxsize=17cm\epsfysize=17.5cm\epsffile{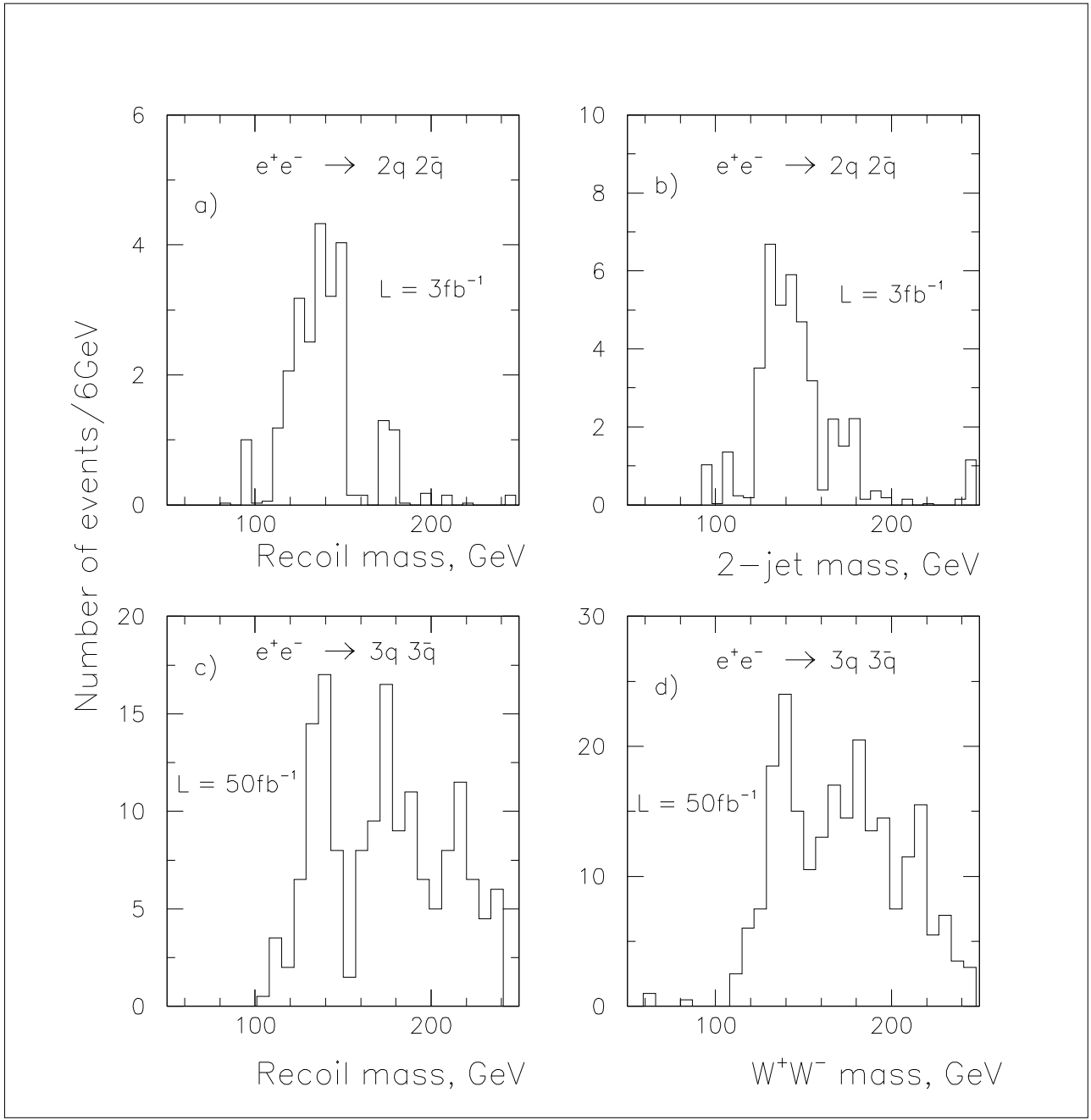}}
\end{center}
\caption[ ]{\sl Recoil mass and 
2-jet  (WW$^*$) mass distributions for the reactions \ee \into 2$q$2$\bar{q}$  \,and 3$q$3$\bar{q}$ with integrated luminosities as indicated.
Background contributions are included.}
\label{fig:4j-6j}
\end{figure*}

Criteria a) - e) remove low-\Pt, low-multiplicity background while
retaining most of the signal events. Cuts f) and g) ensure a 4-jet
topology  with the best Z H interpretation.  The remaining conditions
take care of a good event containment, reasonable jet properties and
b-quark enrichment. After all cuts,  18\% (9\%) of the signal
remains in the 2-jet (recoil) mass distribution.
The smaller efficiency for
the recoil mass is caused by the additional constraint that the mass of
the 2-jet system recoiling against the Higgs has to be within $M_Z \pm$
15 GeV. For an integrated luminosity of 
3 fb$^{-1}$  we expect 23 \hnull events over $\sim$ 5 background events in
the 2-jet mass range $M_H \pm$18 GeV (Fig.5b),  whereas 15 signal
events over $\sim$ 3 background events are observed
in the recoil mass within the range $M_H \pm ^{12}_{30}$ GeV, as seen in
Fig.\ref{fig:4j-6j}a. The surviving background is practically due 
to WW pair production only.

\subsection{The 6-jet channel: \ee \into \znull\hnull \into (\qq)($WW$*)
  \into (\qq)($2q2\bar{q})$}
Due to the large branching fraction of the Higgs into WW*
 expected for $M_H$ =
140 GeV and its sensitivity to discriminate between SM-like
and non SM-like Higgs particles
, it is highly interesting to know which luminosity is needed
to observe a statistically significant \hnull \into WW* signal. The
event selection, however, becomes now more complicated as can be 
anticipated from
the expected topology shown in Fig. 1f.
Here, we consider only hadronic decays  of the vector bosons
taking into account the advantage of their large branching fractions. An
analysis including also the leptonic W \into $ l\nu$ decay may be
found in \cite{bark}, with the result that only a marginal signal could be
extracted. 

As mentioned in sect.4, the absence of a 2-to-6 body event generator
needed for a general analysis analogous to that of sects. 5.1 to 5.4,
implies to proceed as in previous Higgs studies, namely
computing production cross section times branching fractions. Events
of the signal reaction \ee \into \znull\hnull \into (\qq)(WW*)
  \into (\qq)($2q2\bar{q})$ are generated and  confronted with the
expectations from  reactions (1)-(8), (13) and (14) and the 6-jet
 channels \ee \into WWZ, ZZZ
\into $3q3\bar{q}$, which are believed to be the
significant sources of background.

The analysis requires the reconstruction of  6-jet final states
containing four jets from the Higgs and two from the Z.
Basically, the signal can be searched for either in the WW* (\into
$2q2\bar{q}$) invariant mass or in the recoil mass against the
identified \znull. The following criteria have been applied in our
analysis:
\begin{itemize}
\item[a)] the transverse energy of the event should be larger than 50
GeV and below 270 GeV;
\item[b)] the total momentum in beam direction is restricted to $\pm$
80 GeV;
\item[c)]the visible event energy is larger than 220 GeV;
\item[d)]the number of charged particles exceeds 25;
\item[e)]isolated electrons should have an energy smaller than 30 GeV
and a polar angle in the range 12$^0$ to 168$^0$;
\item[f)]the thrust is restricted to be in the range 0.60 to 0.90;
\item[g)]if the event involves 6 jets, the opening angle between any
two of them has to be larger than 20$^0$;
\item[h)]for each jet it is required to have an energy larger than 10
GeV,  $|cos\Theta_{jet}| <$ 0.8 and to involve at least 5 particles;
\item[i)]from all possible Z and H jet pairing only the
combination with smallest $\chi^2$,

\begin{math}
\chi^2 = \frac{(M_Z - M_{ij})^2}{\sigma^2_{M_{ij}}} \: + \: 
\frac{(M_H - M_{klmn})^2}{\sigma^2_{M_{klmn}}} \: + \:
\frac{(M_W - M_{kl})^2}{\sigma^2_{M_{kl}}} ,
\end{math}

is selected. M$_{ij}$ (M$_{klmn}$) is the invariant mass of the jet system
(i,j) (k, l, m, n) and the $\sigma$'s are the corresponding mass
resolutions;
\item[j)]the jet energies are recomputed using energy-momentum
conservation, while keeping their directions at measured values;
\item[k)]the polar angle of the 2-jet pair consistent with the \znull
is restricted to $|$cos$\Theta_{Z}| <$ 0.8;
\item[l)]the event is rejected if more than 4
 tracks with significant impact
parameter, $b_{norm} > 3$, are found (anti-b tag).
\end{itemize}

Low-\Pt, low-multiplicity background is rejected by criteria a) - e), 
while practically all signal events are retained. Once the event has been
selected as a ZH final state by the $\chi ^2$ condition,
cuts for jet containment and good
jet properties are applied. In order to remove background from 
Higgs decays to \bb,
 an anti-b tag is required.

After all cuts applied, 7.5 \% and 6.3 \% of the signal survives in the
4-jet mass respectively recoil mass distribution. The background left is
mainly due to WW pair and $2q2\bar{q}$ production, in the ratio of about
15 : 1. An integrated luminosity of 50 $fb^{-1}$ would allow to observe
32 \hnull events over a background of $\sim$ 18  in the 4-jet mass
distribution, as shown in Fig. 5d. The recoil mass spectrum would
yield a signal of 30 events over $\sim$ 9 background events. The events
counted are within the mass ranges M$_H\pm^{6}_{12}$ GeV respectively
$M_H \pm$12 GeV. Finally, we  notice that the 6-jet
background not accounted for in this study is
expected not to alter our conclusions  significantly.

\subsection{The light quarks and gluon channels: \ee \into \znull\hnull \into
 (\nn)($c\bar{c}$ + gg),
 \ee \into \nn \hnull \into \nn($c\bar{c}$ + gg)
 and \ee \into \znull\hnull \into (\qq)($c\bar{c}$ + gg) \\}

The search for Higgs decays into lighter quarks and gluons takes
advantage of the existence of several production processes. Either the
missing energy channel with Higgs production via the bremsstrahlung or
the fusion mechanism can be searched for or the 4-jet topology with
\znull \into \qq and \hnull \into $c\bar{c}$ + gg decays can be used. Since
the SM branching fraction for 140 GeV Higgs into \cc is only 1.5\%, we have
combined the $c\bar{c}$ and $gg$ decay modes, resulting in a 6.4\% decay
fraction\footnote{It would be very desirable to measure the \hnull
  \into $c\bar{c}$ decay fraction separately. This requires, however, very
  large statistics and an excellent vertex detector to
  collect a relatively clean $c\bar{c}$ event sample}.

\begin{figure*}[h!t]
\begin{center}
\mbox{\epsfxsize=17cm\epsfysize=17.5cm\epsffile{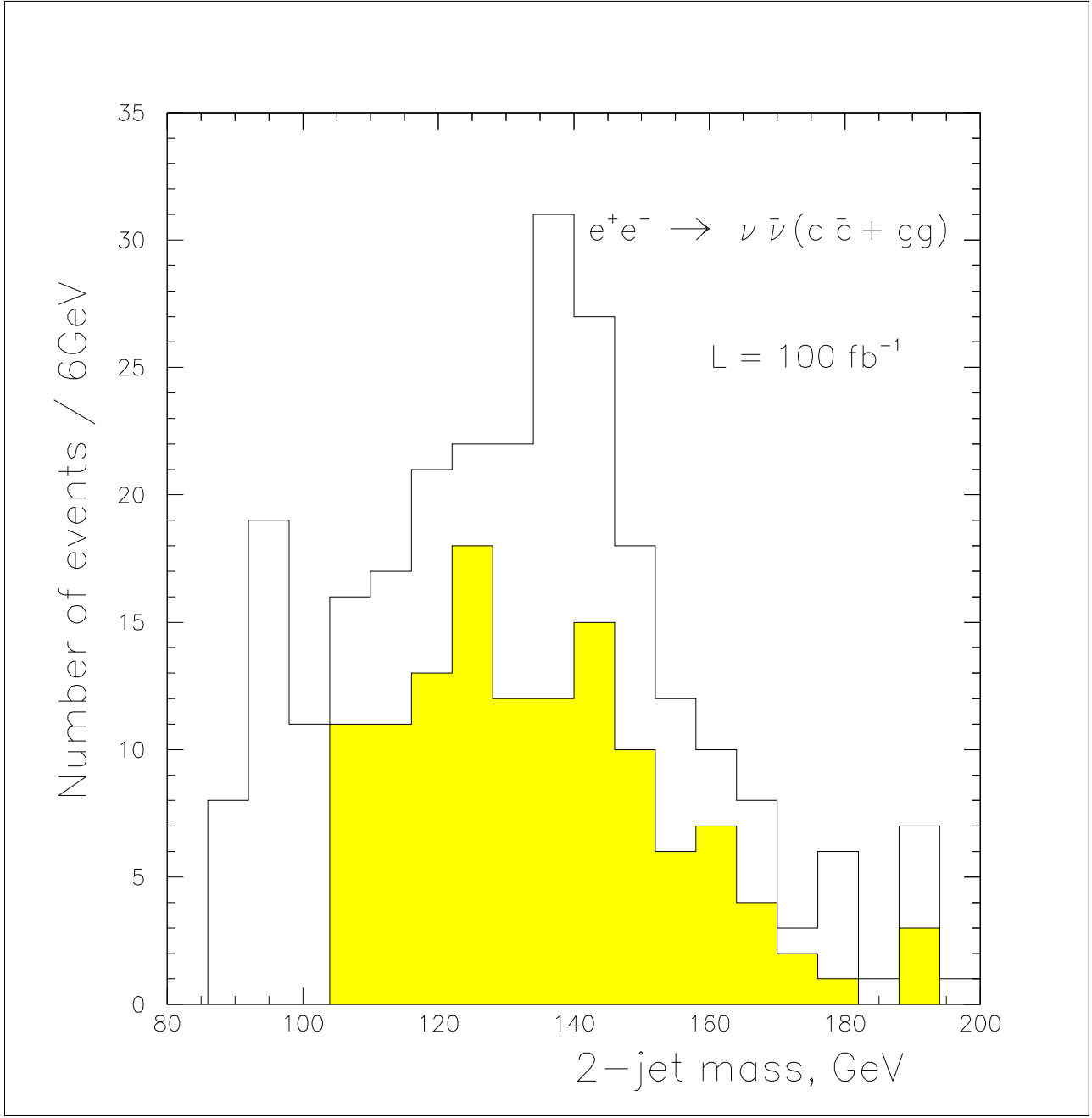}}
\end{center}
\caption[ ]{\sl 2-jet  mass distribution of the reaction \ee \into \nn(\cc + \glgl) for an integrated luminosity of 100 fb$^{-1}$.
The shaded histogram represents the reducible background expected.}
\label{fig:glu-h}
\end{figure*}

The encouraging Higgs discovery potential of the 2-jet missing energy
channel (4) involves to some extent Higgs decays into lighter
quarks, but only at a very small level. Here, we focus on a
a strategy to select \hnull \into $c\bar{c}$ + gg decays
out of the huge background expected. In principle, the 2-jet missing energy
channel should
not considerably suffer from  WW pair or any other background contributions,
because two and only two jets and large missing energy in the
final state are required. However, it has been found that very stringent
selection criteria are needed to distinguish a possible signal
from the remaining background:

\begin{itemize}
\item[a)] the total transverse energy has to be in the range 30 to
  150 GeV;
\item[b)]the total longitudinal momentum is restricted to $\pm$ 120
  GeV;
\item[c)] the visible event energy is larger than 140 GeV and less than
  210 GeV;
\item[d)]isolated electrons with a polar angle between 30$^o$ and
  150$^o$ should have an energy less than 5 GeV;
\item[e)]the number of charged particles per event is larger than 15;
\item[f)]the thrust is required to be below 0.9;
\item[g)]the hadronic system involves two jets with an opening angle
  of at least 80$^o$;
\item[h)]each jet should have an energy larger than 30 GeV,
  $|cos\Theta_{jet}| <$ 0.7 and involves more than 11 particles;
\item[i)] the missing mass has to be consistent with the \znull mass;
\item[j)] the number of tracks with large impact parameter, $b_{norm}
  >$ 3, is less than 4 to suppress \hnull
  \into \bb events (anti-b tag).
\end{itemize}
Fig.\ref{fig:glu-h} shows the combined 2-jet mass distribution for the signal
and the surviving background events,  for 100 fb$^{-1}$
integrated luminosity.
The shaded histogram represents the
reducible background, mainly from WW pair production and, to lesser
extent, from the \tata \qq channel.
 As can be seen, a signal of about 40 \hnull events exists on
$\sim$ 60 events of background within 12 GeV of the Higgs 
mass. Although $S/\sqrt{B}$ is somewhat below 5
 we consider it as convincing and as a demonstration
to observe  \hnull \into $c\bar{c}$ + gg decays.

In the case of the 4-jet event topology, all our attempts to select
a clean \hnull signal were unsuccessful. We were not able to establish
a set of criteria to reject most of the ZZ and WW background and 
more refined constraints have to be
developed in order to make a signal-to-background analysis successful.


\section{Branching fractions of the Higgs boson}
If a Higgs boson is discovered, it is imperative to understand its
nature. Besides its mass the parameters which determine the relation
of the Higgs to the Standard Model are its couplings to vector bosons
and fermions. The reactions studied (see Table 1) allow
to determine the branching fractions for \hnull \into \bb, \tata, WW and
light \qq + gg decays.

Let us consider, as an example, the reaction  \ee \into
\znull\hnull \into $\mu^+\mu^-$ \bb. From the reconstructed \bb
invariant mass the quantity $\;  \sigma(ZH)_{tot} \cdot $  BF(\znull
\into $\mu^+\mu^-)\cdot$  BF(\hnull \into \bb) can be measured. The
branching fraction of the \znull $\:$ 
to  $\mu^+\mu^-$ is well known from
LEP experiments so that in order to obtain
BF (\hnull \into \bb) and its error one has
to  measure the inclusive Higgs bremsstrahlung cross
section $\sigma (ZH)_{tot}$ and to compute

\begin{equation}
BF (H^o \rightarrow b\bar{b}) = \frac{[\sigma(ZH)_{tot} \cdot 
BF (H^o \rightarrow  b\bar{b})]}{\sigma (ZH)_{tot}}
\end{equation}

\subsection{The inclusive cross section $\sigma(ZH)_{tot}$}
The measurement of the inclusive \ee \into \znull\hnull  cross section
is based on the selection of all Z's produced in an \ee  experiment,
with decays into muon and electron pairs. If any $\mu^+\mu^-$ or \ee 
pair detected has a mass consistent with the \znull  mass its recoil mass is
calculated and plotted. From this spectrum  $\sigma$(\ee \into
 \znull\hnull)$_{tot}$ and its error can be estimated taking into
account the dilepton \znull  branching fraction, acceptance
and resolution effects of the detector and the impact of ISR and
beamstrahlung. 

\begin{figure*}[h!t]
\begin{center}
\mbox{\epsfxsize=17cm\epsfysize=17.5cm\epsffile{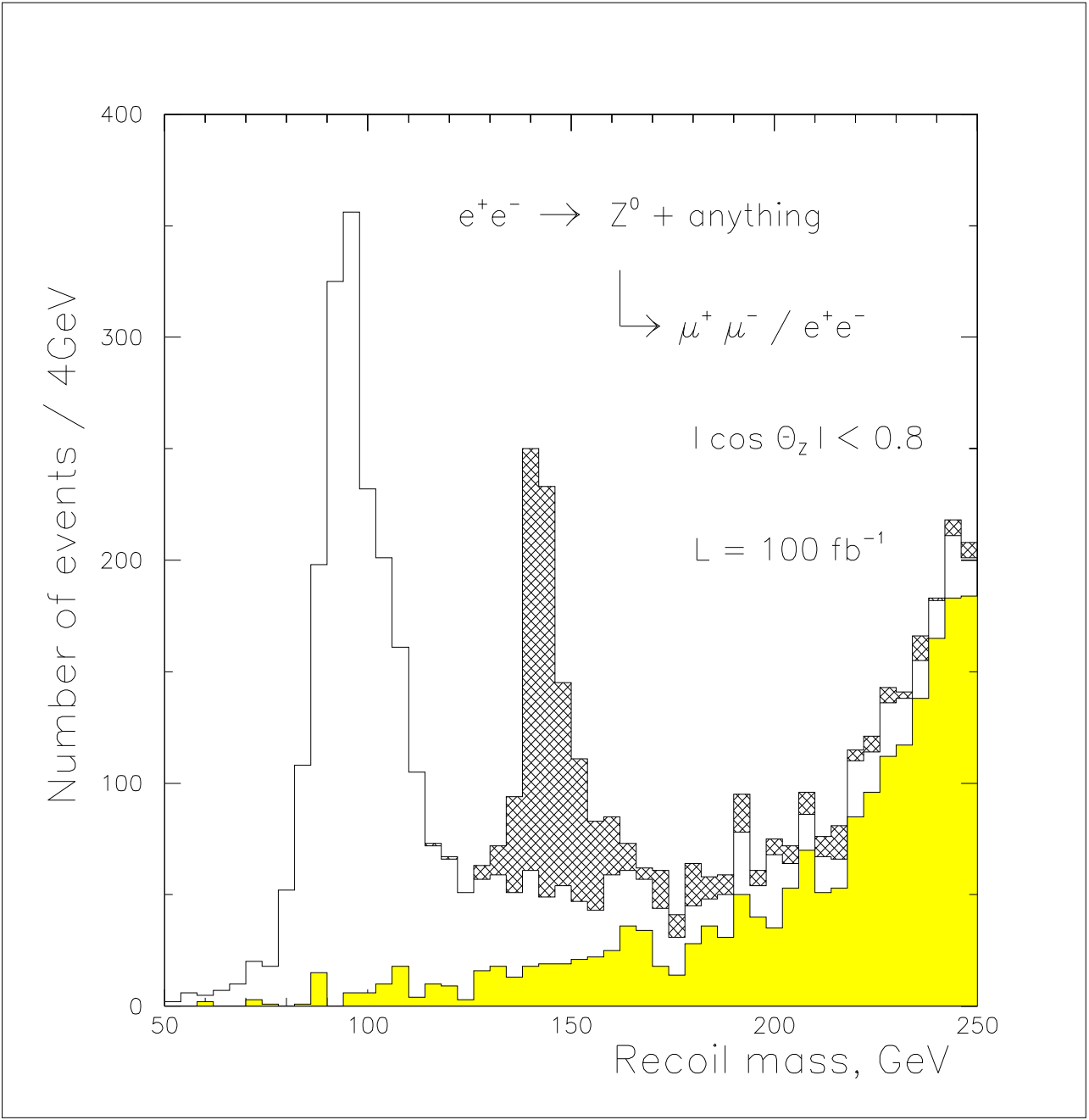}}
\end{center}
\caption[ ]{\sl Inclusive recoil mass distribution in the process \ee \into \znull(\into \mm/\ee) + anything for an integrated luminosity of 100 fb$^{-1}$ and $|cos \Theta_Z| <$ 0.8.
The shaded histogram represents the reducible background expected, whereas the Higgs contribution is shown cross-hatched. The irreducible background for the
\hnull \into WW channel has been neglected.}
\label{fig:inclusive}
\end{figure*}

For all reactions studied so far together with
the additional signal channel \ee \into \znull\hnull \into
($\mu^+\mu^-; e^+e^-$)(WW*) and
allowing for all W decay modes,  we show in Fig.\ref{fig:inclusive} the
recoil mass expected against all
dilepton pairs with M($\mu^+\mu^-)$ =
M$_Z \pm $ 10 GeV and M(\ee) = M$_Z \pm$ 6 GeV and $|cos \Theta_Z| <$
0.8. 
Some improvement of the recoil mass resolution has been achieved by
the constraint M(\lele) = $M_Z$ ~since it is essential that the peak
be as narrow as possible in order to enhance the signal-to-background
ratio. The cut applied on the \znull ~polar angle
is demanded in order to reduce large background from s-channel
Z/$\gamma^*$ (see Fig. 1a) and \znull\znull events. 
The Higgs signal, summed over all its decays,
is clearly visible. It has a non-gaussian shape
which is caused by loss of energy in the initial state due to photon
radiation  and
beamstrahlung resulting to a reduced peak hight
and a long tail on the high mass side.
 The  remaining reducible  background, mainly due to WW pair
and \ee/$\mu^+\mu^-(\gamma)$ production, is shown as hatched histogram
in Fig. 7. In the recoil mass range 138 to 154 GeV we expect 533
\hnull events over $\sim$ 181 background events,
 so that the statistical precision of the inclusive \hnull
signal, $\sqrt{S+B}/S$,
is  $\pm$ 5\%. This accuracy should also hold  for the
HZZ coupling-squared which is proportional to  $\sigma(ZH)_{tot}$.

\subsection{Measurement of the Higgs branching fractions}
In the following we assume the SM couplings of the Higgs particle to
bosons and fermions, a statistical error of 5\%  for the inclusive Higgs
production cross section  $\sigma(ZH)_{tot}$ and determine the
statistical error of the branching fraction 
 BF(\hnull \into xx), for each
of the possible \hnull \into xx decay mode.

The branching fraction~ BF(H \into \bb) can be determined from three
processes, \ee \into \lele \bb, \nn \bb and \qq\bb. Because of the
different statistics and systematics 
expected in these channels, we measure this
fraction for each reaction separately and then combine the results.

The process \ee \into \mm ;\ee \bb $\:$ has the advantage of a simple
signal extraction  with relatively small
background. The selection criteria applied (see
sect.5.1) allow however for some \hnull \into $c\bar{c}$ or other \hnull
decays with badly measured tracks. We found 13 non-\bb
events and subtracted them from the observed signal
in the 2-jet invariant mass resulting to 86 \hnull \into \bb  events over
18 background events, within the range $M_H\pm^{10}_{15}$
GeV. Hence, a 12 \%  statistical error for $\sigma (ZH)_{tot} \cdot$ BF(H
$\rightarrow$ \bb) is obtained from this topology.

The reaction \ee \into \nn \bb benefits from the three times larger
branching fraction of the \znull into \nn ~ w.r.t. muons or electrons
and from a topology (Fig. 1c) which is very distinct from many
important 
background reactions.  Most of the remaining
reducible background is due to WW pair and \ee \qq production, where
the $e^+$ and $e^-$ are lost down the beam pipe. However, the
background is manageable after application of cuts as proposed in
sect.5.3. Taking into account the 16 non-\bb \hnull decays
expected  and demanding the
invisible mass to be close to $M_Z$, we count 460 \hnull \into \bb
events over 95 background events, in the mass range $M_H\pm^{6}_{14}$
GeV. This corresponds to a statistical uncertainty of 5.1 \% for  
$\sigma$(ZH)$_{tot} \cdot$ BF(H $\rightarrow$ \bb).

For the reaction \ee \into $2q2\bar{q}$ $\:$ four well-defined jets are
required in the final state. They have been selected as described in
sect.5.4. In order to reduce the considerable WW and large irreducible  ZZ
contaminations only the jet pairing in best agreement with the
\znull\hnull  hypothesis has been selected. 
Our analysis, after corrections for non-\bb \hnull decays,
results in 479 H$^0$ \into \bb  events over a background of 141
events within $M_H\pm ^{14}_{18}$ GeV, so that a statistical error
of 5.2\% is found
 for  $\sigma$(ZH)$_{tot} \cdot$ BF(\hnull \into \bb).

The total error obtained after combining the three (independent) reactions
must be added in quadrature with the
 uncertainty for the inclusive Higgs
production cross section of 5\%. The resulting statistical error found
for the branching fraction  BF(H $\rightarrow$ \bb) is $\pm$ 6.1\%.

The measurement of the Higgs branching fraction into \tata  turns
out to be more difficult. The reasons are i) a small \hnull \into
\tata decay rate of 3.6 \% , ii) the elusive character of the
neutrinos from $\tau$ decays so that the $\tau$ energies cannot be
directly measured and iii) more background due to missing final state
particle information. Fortunately, since the Higgs is much more
massive then the $\tau$, the $\tau$'s in the \hnull \into \tata decay
receive a tremendous boost, so that, to a good approximation, the
decay products of the tau's fall into a very narrow cone around its
initial direction. A simple rescaling of the $\tau$ energies using
energy-momentum conservation while keeping their direction at measured
values, results, together with the selection criteria of sect.5.2, in
 a clear \hnull \into \tata signal over acceptable background,
as seen in Fig. 3d for an accumulated luminosity of 80 fb$^{-1}$.
For L=100  fb$^{-1}$, we estimate 32
signal events over a background of 15 events, within 12 GeV of the
Higgs mass. Hence, the statistical error expected
for  $\sigma (ZH)_{tot} \cdot$ BF(\hnull \into \tata) would be 21.4 \%,
which, together with the uncertainty of $\sigma (ZH)_{tot}$, yields an
accuracy of  $\pm$ 22 \% for the branching fraction BF(\hnull
\into \tata).

The determination of the branching fraction H \into WW* requires (in
our case) the reconstruction of six jets in the final state, with four
jets from the Higgs  and two from the \znull. The application of
the selection criteria of sect. 5.5 to signal and
 background reactions yields 46 signal events over 44 background
events, in the mass range  $M_H\pm^{14}_{18}$ GeV. After
 corrections 
 for non-WW Higgs decays, we obtain 20.6\% for the statistical
error of $\sigma (ZH)_{tot} \cdot$ BF  (\hnull \into
WW). If this number is added in quadrature with the
error for  $\sigma (ZH)_{tot}$, a $\pm$ 22 \% statistical
uncertainty is measured for the branching fraction of the Higgs into WW.

As has been shown in sect.5.6, it seems possible to extract \hnull
\into $c\bar{c}$ + gg events out of the 2-jet missing
energy event sample. If, within 12 GeV of the Higgs mass, the number of
signal events of Fig.6 are corrected for 4 \hnull \into \bb~events
expected, we obtain~37 signal events on a background of~61 events. This
yields, after convolution with the uncertainty of  $\sigma
(ZH)_{tot}$, a $\pm$ 28\% statistical error for the
branching fraction
BF(\hnull \into $c\bar{c} + gg$).


\section{Summary}
In this study we have considered the physics potential of the NLC for
the Higgs sector, assuming the validity of the Standard Model. In order to make a
signal-to-background analysis as meaningful as possible, we have
consistently evaluated both signal and background rates at leading
order. Our simulation studies include

\begin{itemize}
\item the complete matrix elements for the 2-to-4 body processes \ee
  \into $2f2\bar{f}$ beyond the usual factorization approximation;
\item all import SM Higgs decay channels, except H \into ZZ*;
\item initial state radiation and beamstrahlung;
\item a detector response, with parameters of the detector as recently
  proposed in the 'Conceptual Design Report' for the S-band and TESLA
  linear collider options;
\item all reducible background reactions expected to contribute.
\end{itemize}

We demonstrate that a Higgs boson with a mass of 140 GeV can be easily
discovered in the reaction \ee \into $\nu\nu$\qq within a few days of
running of a 360 GeV collider delivering a luminosity of $5 \cdot
10^{33} cm^{-2} sec^{-1}$. Soon after its discovery, the 4-fermion
channel ~\ee \into $2q2\bar{q}$ ~would establish the Higgs despite of
large expected background due to ZZ and WW pair production. The
usually considered gold-plated Higgs discovery channel \ee \into \mm; \ee
\qq provides a clean event sample and allows for simple
\hnull selection procedures. However, due to the small \znull \into
\lele
~branching fraction an integrated luminosity of about 10 fb$^{-1}$ is
needed to observe a convincing signal. This reaction has the great
advantage of being independent of the Higgs decay modes when the
recoil mass technique is used. Furthermore,
 from the recoil mass spectrum the
inclusive Higgs production cross section can be
measured with  5\% uncertainty for an integrated luminosity of 100 fb$^{-1}$
which in turn allows to determine the HZZ coupling with high precision.
 Other
decay modes of the Higgs, such as \hnull \into WW or \tata, are 
unlikely to be useful for its detection. An integrated luminosity of about
50 fb$^{-1}$ (or more) appears necessary to observe statistically
significant signals in these channels. We have, to our knowledge
for the first time, clearly demonstrated the capability of an \ee collider to
detect Higgs decays into lighter quarks and gluons.

For an integrated luminosity of 100 fb$^{-1}$, we have determined the
statistical errors of the Higgs branching fractions to \bb, \tata,
WW$^{(*)}$ and to $c\bar{c} + gg$. The results obtained,
convoluted with an  5\% error for the inclusive Higgs
production cross section, are summarized in Table \ref{tab:3}. 
\begin{table}[h!t]
\begin{center}
\begin{tabular}{|l|l|}
\hline
& \\
Branching fraction & Expected error \\
& \\
\hline
& \\
BF (\hnull \into \bb) & $\pm$ 6.1\% \\
& \\
BF (\hnull \into \tata) & $\pm$ 22\% \\
& \\
BF (\hnull \into WW*) & $\pm$ 22\% \\
& \\
BF (\hnull \into $c\bar{c} + gg$) & $\pm$ 28\% \\
& \\
\hline
\end{tabular}
\end{center}
\caption[ ] {\sl Numerical values of the branching fraction errors
  assuming SM couplings for the Higgs boson with a mass of 140 GeV, a
  cm energy of 360 GeV and an integrated luminosity of 100 fb$^{-1}$.}
\label{tab:3}
\end{table}
As can be seen,
many of the Higgs decay modes are accessible to experimental
consideration and most of the measurements would have an error around
 22\% or better so that a powerful discrimination between the SM-like
Higgs or SUSY-like Higgs can be achieved, even with our relatively
simple analysis procedures.

We would like to emphasize once more that for the production and decay
channels considered in this paper,
only three-level calculations were made and that
the selection criteria proposed are not yet optimized such that significant
improvements can be expected once the precise detector behaviour
is taken into account.

\section*{Acknowledgments}

We would like to thank our colleagues within the course of the
Conceptual Design Report for many discussions. A.Pukhov is
acknowledged for his support and help with CompHEP calculations
and D.Schulte for providing the TESLA beamstrahlung
parameters. S.Sh. also thanks the DESY-Zeuthen TESLA group for the
kind hospitality, in particular P.S\"{o}ding for his interest and support.


\end{document}